\RequirePackage[2020-02-02]{latexrelease}
\documentclass[%
 aip,pof
 amsmath,amssymb,
 reprint,%
]{revtex4-1}

\usepackage{amsmath}
\usepackage{graphicx}
\usepackage{dcolumn}
\usepackage{bm}

\usepackage{caption,subcaption}

\usepackage[utf8]{inputenc}
\usepackage[T1]{fontenc}
\usepackage{mathptmx}

\graphicspath{{figures/}}

\begin{document}

\preprint{AIP/123-QED}

\title{Numerical simulation of solitary gravity waves on deep water
  with constant vorticity}

\author{A. S. Dosaev}
  \email[Author to whom correspondence should be addressed: ]{dosaev@ipfran.ru}
  \affiliation{Nonlinear Geophysical Processes Department,
    Institute of Applied Physics, Nizhny Novgorod, Russia}

\author{M. I. Shishina}
  \affiliation{Nizhny Novgorod Planetarium n.a. G.~M.~Grechko,
  Nizhny Novgorod, Russia}

\author{Yu. I. Troitskaya}
  \affiliation{Nonlinear Geophysical Processes Department,
    Institute of Applied Physics, Nizhny Novgorod, Russia}

\date{\today}

\begin{abstract}
  We present a numerical study of essentially nonlinear dynamics of
  surface gravity waves on deep water with constant vorticity using
  governing equations in conformal coordinates. The dispersion
  relation of surface gravity waves on shear flow is known to have two
  branches, one of which is weakly dispersive for long waves. Weakly
  nonlinear evolution of the waves of this branch can be described by
  the Benjamin-Ono equation, which is integrable and has soliton and
  multi-soliton solutions. Currently, the extent to which the
  properties of such solitary waves obtained within the weakly
  nonlinear model are preserved in the exact Euler equations is
  unknown. We investigate the behaviour of this class of solitary
  waves without the restrictive assumption of weak nonlinearity by
  using the exact Euler equations. The evolution of localized initial
  perturbations leading to the formation of single or multiple
  solitary waves is modeled, and the properties of finite-amplitude
  solitary waves are discussed. We show that within the framework of
  the exact equations, two-soliton collisions are almost elastic, but
  in contrast to solutions of the Benjamin-Ono equation the waves
  receive a phase shift as a result of the interaction.
\end{abstract}

\maketitle

\section{\label{sec:level1}Introduction}

Waves in oceans or inland basins often propagate on vertically sheared
currents, their dynamics being significantly affected by wave-current
interaction \cite{peregrine1976interaction}. On vertically sheared
currents, which may be produced by wind stress or bottom friction, the
nonlinear properties of waves are modified, such as the rate of growth
of modulational instability \cite{thomas2012nonlinear} or breaking
conditions \cite{banner1974incipient}. In particular, using
perturbational analysis, Shrira observed \cite{shrira1986nonlinear}
that shear flows support a specific kind of solitary gravity wave that
is absent in deep water without shear.

A special case, one for which theoretical treatment can be greatly
simplified, is that of a two-dimensional flow with uniform shear
(constant vorticity) because all perturbations to the velocity field
are then strictly potential, as follows from Kelvin's circulation
theorem. In this setting, many results concerning periodic and
solitary stationary waves have been obtained numerically using the
boundary integral method
\cite{simmen1985steady,dasilva1988steep,pullin1988finite}; the
parameter space study was later extended by Vanden-Broeck
\cite{vanden1994steep,vanden1996periodic} to find more solution
families. (See also a more recent study of periodic waves by
S.~Dyachenko and Hur \cite{dyachenko2019stokes,dyachenko2019folds} that
uses a conformal mapping technique.)

The scope of the present paper is confined to deep water waves on a
uniform shear flow propagating in the direction of the shear (negative
vorticity in our notation, see Section~\ref{sec:basic}). Such waves become
weakly dispersive in the long wavelength limit, and their weakly
nonlinear evolution is described by the Benjamin-Ono (BO) equation
\cite{shrira1986nonlinear}. The BO equation is integrable and has
stable soliton solutions; the solitons of the BO equation interact
elastically and do not undergo a phase shift after the interaction
\cite{matsuno1979exact,matsuno1980interaction}. Our aim is to examine
the behaviour of finite-amplitude solitary waves and to determine the
extent to which they retain the properties of their weakly-nonlinear
counterparts. In contrast to the works on finite-amplitude waves cited
above, we do not limit our study to stationary solutions; we use
evolution equations to model solitary waves formation from a localized
initial perturbation (which also ensures stability of the obtained
waves), as well as interaction of two solitary waves.

The simulations are performed within the framework of the full Euler
equations for deep water waves with constant vorticity. The governing
equations in conformal coordinates for a constant vorticity flow with
a free surface over an arbitrary bottom profile were obtained by Ruban
\cite{ruban2008explicit} and were independently obtained for a flat
bottom by Choi \cite{choi2009nonlinear}. Here we use a slightly
different formulation that employs Dyachenko variables
\cite{zakharov2002new} to improve numerical stability of the system;
the corresponding governing equations were derived by Dosaev et
al. \cite{dosaev2017simulation}

Of high relevance to solitary waves modeling is also a recent
development of the conformal mapping methodology related to the use of
auxiliary conformal coordinates with adaptive resolution.  Tanaka
\cite{tanaka1983stability} proposed a change in independent variables
that allows to increase spatial resolution of the conformal grid in a
neighborhood of a selected point (e.g., on the wave crest). Adaptive
conformal coordinates were used to study linear stability of gravity
waves \cite{tanaka1983stability,longuet1997crest}, including waves
with constant vorticity \cite{murashige2020stability}, and to obtain
solutions with very steep profiles in some of the mentioned works on
stationary waves \cite{dyachenko2019stokes,dyachenko2019folds}. A
detailed study by Lushnikov et al. \cite{lushnikov2017new} has
revealed that Hilbert transform can be computed on the adaptive grid
as efficiently as it was in the original representation; the technique
therefore integrates very naturally into the conventional conformal
mapping approach. In the present work we employ adaptive coordinates
to speed up stationary profile computation. Although applying the
change in variables to the evolution equations is straightforward,
too, a method for resolving arbitrary number of sharp profile
features, as well as a robust strategy for dynamically adjusting the
resolution during a simulation, still remains to be developed.

The paper is organized as follows. In Section~\ref{sec:basic}, we
describe the physical model and introduce governing equations in
conformal coordinates that we use in our simulations. In
Section~\ref{sec:stwave}, stationary wave profiles are obtained and
their characteristics are compared with BO solutions. The results of
numerical simulations are presented in Section~\ref{sec:sim}: we model
formation of solitary waves from disintegration of a localized
perturbation, and study collision of two solitary waves using the
profiles obtained in Section \ref{sec:stwave}. Section
\ref{sec:conclusion} summarizes the main results of the work.

\section{\label{sec:basic}Basic equations}

\subsection{Governing equations in Cartesian coordinates}

Consider a two-dimensional flow of an ideal incompressible fluid with
a free surface. In the $(x,y)$-plane, with the $y$-axis directed
upward and the $x$-axis coinciding with the fluid surface at rest, we
describe motion of the fluid using a stream function~$\Psi$, which is
connected to velocity components $v_x,v_y$ as
\begin{displaymath}
  v_x=\frac{\partial\Psi}{\partial y},\quad
  v_y=-\frac{\partial\Psi}{\partial x}.
\end{displaymath}
For waves propagating on a shear current with uniform vorticity
distribution ($\omega=-\Delta\Psi=const$), the stream function can be
decomposed into a sum
\begin{displaymath}
  \Psi(x,y,t) = -\frac{\omega y^2}2 + \psi(x,y,t),
\end{displaymath}
where the first term, which corresponds to the rotational component of the
velocity field, is time-independent and the perturbational part
$\psi$ satisfies $\Delta\psi=0$.

The harmonic conjugate to~$\psi$, which we denote~$\varphi$,
is defined by the Cauchy-Riemann conditions
\begin{displaymath}
  \frac{\partial\varphi}{\partial x}=\frac{\partial\psi}{\partial y},\quad
  \frac{\partial\varphi}{\partial y}=-\frac{\partial\psi}{\partial x},
\end{displaymath}
and is the potential for the same velocity perturbation defined by the
stream function~$\psi$:
\begin{displaymath}
  v_x=-\omega y + \frac{\partial\varphi}{\partial x},\quad
  v_y=\frac{\partial\varphi}{\partial y}.
\end{displaymath}
Together $\varphi$ and $\psi$ form a ``complex potential'',
$\theta=\varphi+i\psi$, which is an analytic function of a complex
variable~$z=x+iy$ in the flow domain.

At the free surface~$y=f(x,t)$, kinematic and dynamic boundary
conditions must be satisfied:
\begin{equation}\label{eq:kin_bc}
  \frac{\partial f}{\partial t}
  + \left(-\omega f+\frac{\partial\varphi}{\partial x}\right) \frac{\partial f}{\partial x}
  - \frac{\partial\varphi}{\partial y} = 0,
\end{equation}
\begin{equation}\label{eq:dyn_bc}
  \frac{\partial\varphi}{\partial t}
  + \frac 1 2 |\nabla\varphi-\omega y\mathbf{x}_0|^2
  + \omega\psi - \frac{\omega^2y^2}{2} + \frac{p_a}{\rho} + gy = 0,
\end{equation}
where~$\mathbf{x}_0$ is a unit vector in the direction of the
$x$-axis, $g$~is the gravity acceleration, and $p_a$~is pressure at the
free surface. Inside the flow domain, $\varphi$ must satisfy the
Laplace equation
\begin{equation}\label{eq:laplace_phi}
  \Delta\varphi=0.
\end{equation}
We confine our study to deep water waves, thus assuming a boundary
condition at infinity:
\begin{equation}\label{eq:inf_bc}
  |\nabla\varphi| \to 0 \quad\textrm{as}\quad y\to -\infty.
\end{equation}
Set (\ref{eq:kin_bc})--(\ref{eq:inf_bc}) is equivalent to exact
Euler equations and completely defines dynamics of the system.

For small amplitude waves propagating in the positive direction of the
$x$-axis, the system (\ref{eq:kin_bc})--(\ref{eq:inf_bc}) gives the
following dispersion relation:
\begin{equation}\label{eq:lin_disp}
  c^2k - \omega c - g = 0,
\end{equation}
where $k$ is the wavenumber and $c$ is the phase velocity. The parameters
of the system, $g$ and $\omega$, have only one combination with
dimension of length
\begin{displaymath}
  \lambda_g=g/\omega^2,
\end{displaymath}
which is therefore a natural length scale for the problem. According
to~(\ref{eq:lin_disp}), long waves ($k\lambda_g \ll 1$) propagating in
the direction of shear (i.e. in the positive direction of the
$x$-axis, if~$\omega < 0$) are weakly dispersive:
\begin{equation}
  c \approx \frac{g}{|\omega|} - \frac{gk}{|\omega|^3} = c_0(1-k\lambda_g),
\end{equation}
where $c_0=g/|\omega|$ is the limiting phase velocity for
small-amplitude long waves. The weakly nonlinear evolution of such
long waves is described by the Benjamin-Ono equation
\cite{shrira1986nonlinear}
\begin{equation}\label{eq:bo_eqn}
  \left(\frac{\partial}{\partial t} + c_0\frac{\partial}{\partial x}\right) f
  - \omega f\frac{\partial f}{\partial x}
  - \frac{g^2}{\omega^3}\hat{H}\frac{\partial^2 f}{\partial x^2}=0,
\end{equation}
where $\hat{H}$ is Hilbert transform:
\begin{displaymath}
  \hat{H}f(x) = \frac 1 \pi P.V.\int_{-\infty}^{+\infty}\frac{f(\chi)d\chi}{\chi-x}.
\end{displaymath}

\subsection{Governing equations in conformal coordinates}

Let us map the flow domain onto a lower half-plane of complex variable
$\zeta=\xi+i\eta$ via a (time-dependent) conformal mapping
\begin{displaymath}
  x+iy=z(\zeta,t) = \xi + i\eta + \tilde x(\xi,\eta,t) + i\tilde y(\xi,\eta,t).
\end{displaymath}
The free surface \mbox{$y=f(x,t)$} is thus mapped onto the real
axis~$\eta=0$.

The resulting parameterization of the free surface is particularly
convenient because in the new coordinates~$(\xi,\eta)$ functions
$\varphi$ and $\psi$ remain harmonic, and their values on the free
surface $\eta=0$ are related through the Hilbert transform
\begin{displaymath}
  \psi(\xi,\ldots)=\hat{H}\varphi(\xi,\ldots).
\end{displaymath}

In conformal coordinates, boundary conditions~(\ref{eq:kin_bc}) and
(\ref{eq:dyn_bc}) at the free surface~$\eta=0$ become
\cite{choi2009nonlinear,shishina2016}
\begin{equation}\label{eq:conf_xy}
  x_\xi y_t - x_ty_\xi = -\hat{H}\varphi_\xi + \omega yy_\xi,
\end{equation}
\begin{equation}\label{eq:conf_phi}
  \begin{split}
    \varphi_t + \varphi_\xi\hat{H}\left[\frac{-\psi_\xi+\omega yy_\xi}{J}\right]
    -\frac{\omega yx_\xi}{J}\varphi_\xi \\
    +\frac{\varphi_\xi^2-\psi_\xi^2}{2J} + \omega\psi + \frac{p_a}\rho + gy &= 0,
  \end{split}
\end{equation}
where $J=|z_\xi|^2$. They can also be written in terms of analytical
functions~$z$ and $\theta$ as
\begin{equation}\label{eq:zt}
  z_t = iUz'
\end{equation}
\begin{equation}\label{eq:thetat}
  \begin{split}
    \theta_t &= iU\theta' - (1+i\hat H)(
    \frac{|\theta'|^2}{2|z'|^2} - \omega y\ \mathrm{Re}\frac{\theta'}{z'} + \frac{p_a}\rho
    )\\
    &+ i\omega\theta + ig(z-\xi),
  \end{split}
\end{equation}
where prime denotes differentiation with regard to~$\zeta$, and
\begin{displaymath}
  U = (1+i\hat{H})\left[\frac{-\psi_\xi+\omega yy_\xi}{J}\right].
\end{displaymath}

For the purpose of modeling the evolution of the system, the
equations~(\ref{eq:zt}) and (\ref{eq:thetat}) can be rewritten in a form,
that is more suitable for numerical integration, through a change of
variables
\begin{equation}\label{eq:dyach_def}
  R=\frac{1}{z'},\quad V=\frac{i\theta'}{z'}.
\end{equation}
The resulting governing equations are as follows
\cite{dosaev2017simulation}:
\begin{eqnarray}
  \label{eq:rt}
  R_t &=& i(UR'-U'R)\\
  \label{eq:vt}
  V_t &=& i\left(UV'-R\hat{P}'(|V|^2-2\omega\ \mathrm{Im}\ z\ \mathrm{Im}\ V+\frac{2p_a}{\rho})\right) \nonumber\\
      && + g(R-1) + i\omega V,
\end{eqnarray}
where $\hat{P}=\frac 1 2 (1+i\hat{H})$ is a projection operator, and
\begin{displaymath}
  U = \hat{P}(RV^*+R^*V-2\omega\ \mathrm{Im}\ z\ \mathrm{Im}\ R).
\end{displaymath}
When $\omega=0$, system (\ref{eq:rt}) and (\ref{eq:vt}) is reduced to
the Dyachenko equations \cite{zakharov2002new}. In the rest of this
work the external pressure $p_a$ is assumed to be zero.

Derivation of the BO equation in conformal coordinates is
given in the Appendix.

\section{\label{sec:stwave}Finite-amplitude stationary waves}

In this section we obtain periodic stationary solutions of the exact
equations of motion and discuss what new properties they possess
compared to the weakly nonlinear BO model. Our interest to the
periodic solutions in the context of solitary waves stems from the
fact that BO solitons are algebraic with $f \sim x^{-2}$ tails; they,
therefore, cannot be represented with reasonable precision on a
periodic coordinate grid that our numerical scheme (based on fast
Fourier transform) utilizes. As a result, we must limit ourselves to
studying periodic solutions, approaching solitary waves through
increase in spatial period.


Periodic solution of the BO equation takes the form
\cite{benjamin1967internal}
\begin{equation}
    \begin{split}
      f(x,t) = f_0 + \frac{akd}{4i} \Big(
        &\cot \frac{k}{2}(x-ct-id)\\
        &-\cot \frac{k}{2}(x-ct+id) \Big),
    \end{split}
\end{equation}
where $k=2\pi/L$ is the wavenumber and the parameters satisfy
\begin{equation}\label{eq:bo_ad}
  ad=4\lambda_g^2,
\end{equation}
\begin{displaymath}
  c = c_0 + |\omega| f_0 + \frac{|\omega|a}{4} kd \coth kd.
\end{displaymath}
The crest becomes more localized as the wave ``amplitude''~$a$
increases, and in the limit $kd \to 0$ the profile becomes Lorentzian:
\begin{equation}\label{eq:bo_lorentz}
  f(x,t) = \frac{a}{1+(x-ct)^2/d^2},
\end{equation}
\begin{equation}
  \label{eq:bo_c}
  c=c_0 + \frac{|\omega| a} 4.
\end{equation}


In order to obtain an equation for the exact stationary wave profile
in conformal coordinates we seek solutions
for~(\ref{eq:conf_xy}) and (\ref{eq:conf_phi}) in the form
\begin{displaymath}
  y=y(\xi-ct),\quad
  x=\xi+\tilde{x}(\xi-ct),\quad
  \varphi=\tilde\varphi(\xi-ct)-gb_0t.
\end{displaymath}
Substitution into (\ref{eq:conf_xy}) gives
\begin{equation}\label{eq:stat_kin_bc}
  (c+\omega y)y_\xi=\hat{H}\varphi_\xi.
\end{equation}
Using (\ref{eq:stat_kin_bc}), we eliminate potential~$\varphi$ from
(\ref{eq:conf_phi}) and obtain an equation for the wave profile in
conformal parameterization~$z(\xi)$:
\begin{equation}\label{eq:stat_profile}
  gy-gb_0=\frac{c^2}2 -
  \frac{\left(c+\omega(yx_\xi+\hat{H}(yy_\xi))\right)^2}{2J}.
\end{equation}

Equations of motion (\ref{eq:kin_bc})--(\ref{eq:inf_bc}) contain only
two parameters of the medium, that is, $g$ and $\omega$, and exactly
two parameters we can exclude from the equations by choosing
appropriate units of measurement for length and time. This means that
the nondimensionalized form of the equations obtained by converting to
nondimensionalized coordinates and time
\begin{displaymath}
  X=x/\lambda_g,\quad Y=y/\lambda_g,\quad
  \Xi=\xi/\lambda_g,\quad T=|\omega|t
\end{displaymath}
will not contain any parameters that depend on the medium, and a
similarity law can be established for motions of the system at various
$(g,\omega)$. Therefore, profiles of stationary waves that for
$\omega<0$ satisfy a nondimensionalized equation
\begin{displaymath}
  Y + \frac{1}{2J}\left(C-YX_{\Xi}-\hat{H}(YY_{\Xi})\right)^2=const
\end{displaymath}
are universal and only depend on dimensionless phase velocity
$C=c/c_0$ and wavelength $\Lambda=L/\lambda_g$. Solitary waves,
which can be considered a special case of a stationary wave with
infinite wavelength, constitute a one-parameter family of solutions,
and all of their dimensionless parameters (such as wave height
$h/\lambda_g$ or phase velocity $c/c_0$) can be found as functions of
a single parameter.

We seek a solution to (\ref{eq:stat_profile}) in a form of a pole
expansion
\begin{equation}\label{eq:pole_sum}
  z(\zeta) - \zeta = z_0 + \sum_n a_n\frac{kd_n}{2} \cot \frac{k}{2}(\zeta-id_n).
\end{equation}
where the constant $z_0=iy_0$ is chosen so that the mean water level
remains zero:
\begin{displaymath}
  \int yx_\xi d\xi = 0.
\end{displaymath}
If all $d_n>0$, then $z(\zeta)$ only has poles in the upper
half-plane, and is analytical in the flow domain
$\mathrm{Im}\ \zeta<0$.  For a given phase velocity $c$ we can obtain
the optimal set of coefficients $\{a_n,d_n\}$ by minimizing the
residual of the stationary wave equation (\ref{eq:stat_profile}):
\begin{displaymath}
  r(a_1,\ldots,d_1,\ldots) = gy + \frac{\left(c+\omega(yx_{\xi}+\hat{H}(yy_{\xi}))\right)^2}{2J}
  - \frac{c^2}{2},
\end{displaymath}
\begin{displaymath}
  \int |r|^2 du \to min.
\end{displaymath}
We use fast Fourier transform to compute the derivatives and Hilbert
transform. Number of points of the spatial grid required for the
computation of the residual can be greatly reduced by utilizing an
adaptive conformal grid \cite{lushnikov2017new} with increased
resolution in the vicinity of the wave crest. The adaptive grid is
equidistant in an auxiliary coordinate $q$:
\begin{displaymath}
  \frac{\xi}{2} = \arctan\left(\alpha\tan\frac{q}{2}\right).
\end{displaymath}
Hilbert transform, computed in the $q$~coordinate, differs from that
in the $\xi$~coordinate only by a constant, which can be easily found
from the condition that mean value of the transform in the
corresponding coordinate is zero. After obtaining profile
(\ref{eq:pole_sum}), the corresponding velocity field is recovered
from~(\ref{eq:stat_kin_bc}). This stationary solution in form of
$z(\xi)$ and $\varphi(\xi)$ can also be converted to Dyachenko
representation according to (\ref{eq:dyach_def}), which will be used
in Section~\ref{sec:sim} for constructing initial conditions for numerical
simulations.

Using this simple method and by varying spatial period $L$ and phase
velocity $c$, we were able to obtain waves of any height up to the
point of phase velocity extremum (which is located around
$h\approx 0.55\lambda_g$ for long waves $L>10^3\lambda_g$). Here the
wave height is defined as
\begin{displaymath}
  h = \max_x f - \min_x f = \max_\xi y - \min_\xi y.
\end{displaymath}
Less than 10 poles in the expansion (\ref{eq:pole_sum}) were typicaly
required to achieve the best approximation, further minimization of
the residual being obstructed by numerical errors in its computation
(within IEEE 754 double precision). The considered range of amplitudes
contains a maximum of wave energy (see below), which means that only
part of the range represents stable solutions. Since our primary goal
is to study solitary wave interactions, only stable solutions are of
interest in the context of this work.


\begin{figure}
  \includegraphics[width=\linewidth]{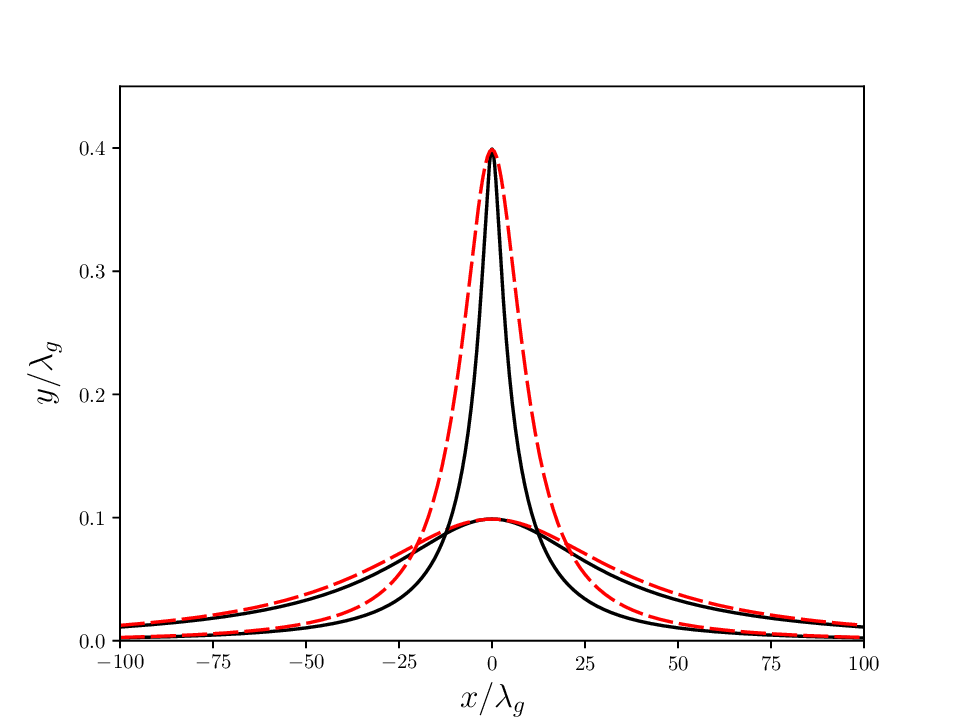}
  \caption{Stationary waves of the exact equations (solid lines) and
    the BO solutions of the same height (dashed lines). Wave heights:
    $h=0.1\lambda_g,\ 0.4\lambda_g$, wavelength $L=10^4\lambda_g$.}
  \label{fig:prof_cmp}
\end{figure}

\begin{figure}
  \centering
  \begin{subfigure}{\linewidth}
    \includegraphics[width=\textwidth]{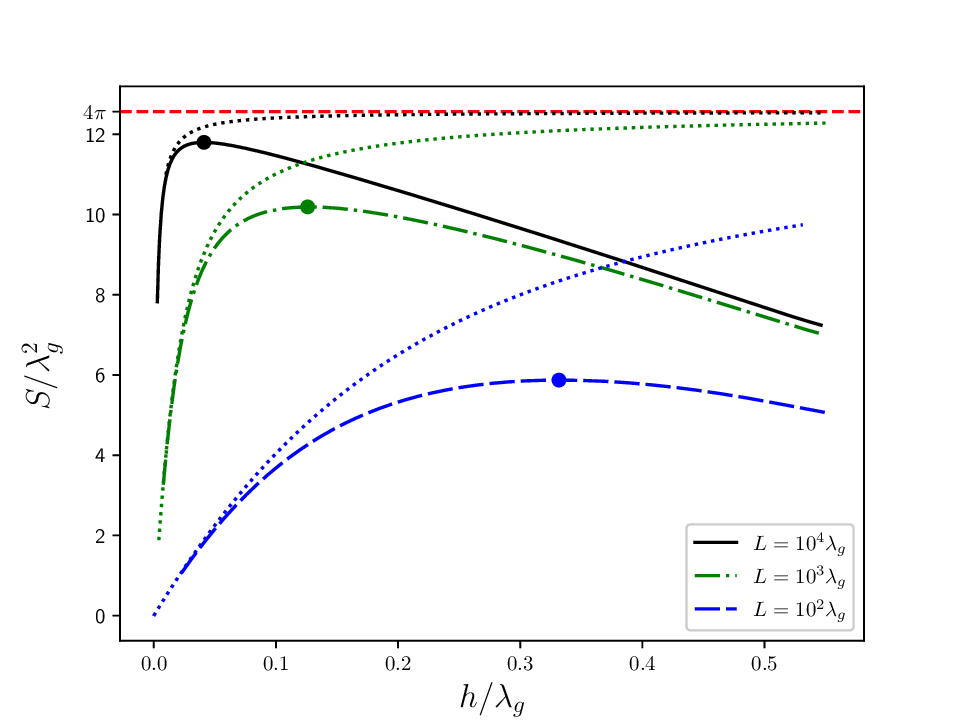}
    \caption{}
    \label{fig:stat_prm:volume}
  \end{subfigure}
  \begin{subfigure}{\linewidth}
    \includegraphics[width=\textwidth]{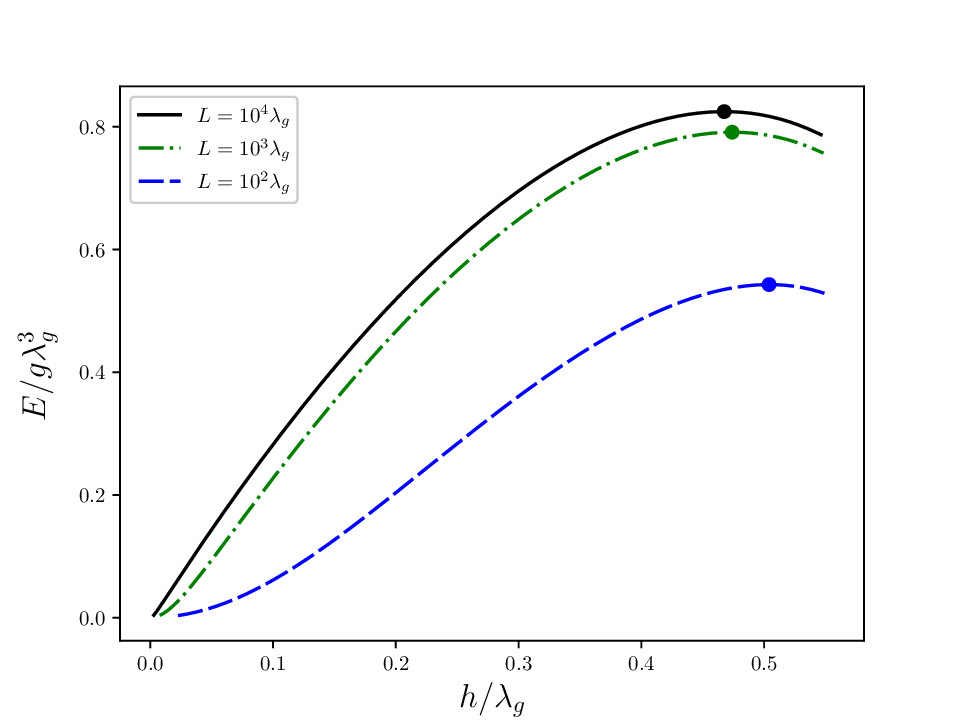}
    \caption{}
    \label{fig:stat_prm:energy}
  \end{subfigure}
  \begin{subfigure}{\linewidth}
    \includegraphics[width=\textwidth]{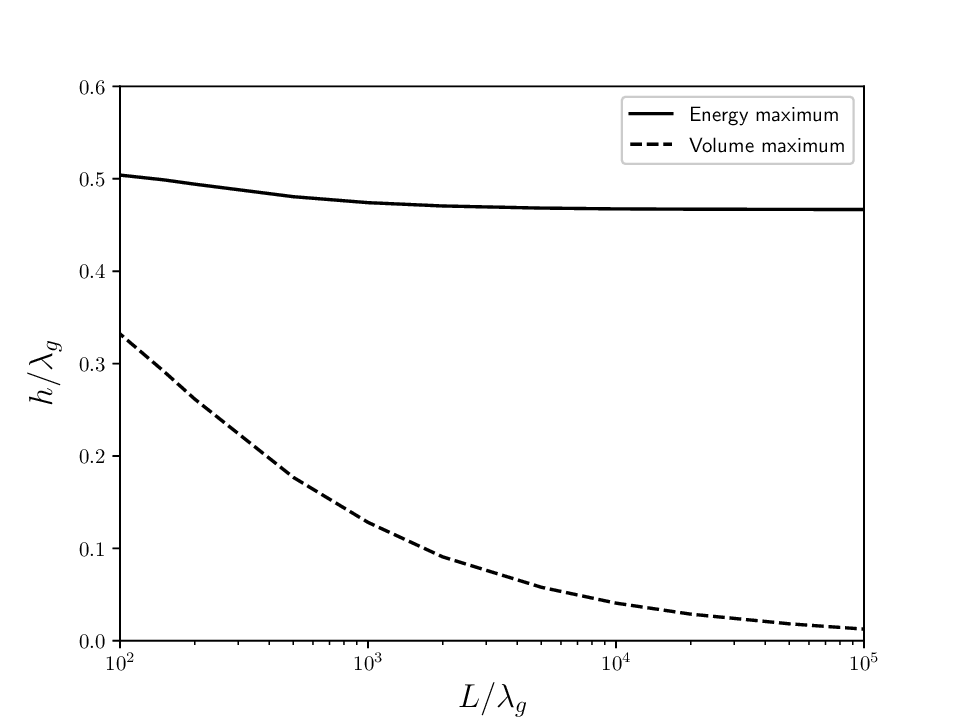}
    \caption{}
    \label{fig:stat_prm:maxima}
  \end{subfigure}
  \caption{(\subref{fig:stat_prm:volume}) Profile volume and
    (\subref{fig:stat_prm:energy}) wave energy as functions of wave
    height; dotted lines represent volume of the BO solutions and markers
    show locations of maxima for solutions of the exact
    equations. (\subref{fig:stat_prm:maxima}) Positions of energy and
    volume maxima as functions of wave period.}
  \label{fig:stat_prm}
\end{figure}

Figure~\ref{fig:prof_cmp} shows example profiles of stationary waves
with heights $h=0.1\lambda_g,\ 0.4\lambda_g$ and wavelength
$L=10^4\lambda_g$, together with periodic BO profiles with the same
parameters. Such long waves already have a great resemblance to
solitons, because for the profiles in the figure the crest widths at
half maximum constitute less than 0.01 of the wave period. One can see
that although the exact and BO solutions are very similar
at~$h=0.1\lambda_g$, for higher waves the crests of the exact profiles
become noticeably more narrow and sharp than the crests of their BO
counterparts, despite both profiles having very close asymptotics at
large distances from the crest.

To better illustrate this observation we introduce profile ``volume'',
which we define as:
\begin{displaymath}
  S = \int_{-L/2}^{L/2}f(x) dx - L\min_x f(x)
\end{displaymath}
According to~(\ref{eq:bo_ad}), the volume of BO solitons (as well as
very high periodic waves) does not depend on wave amplitude and is
equal to~$S_{BO}=4\pi\lambda_g^2$. Profile volume $S$ as a function of
wave height~$h$ for various wave periods $L$ is plotted in
figure~\ref{fig:stat_prm:volume}; as the figure shows, exact profiles
do not just have a different shape of their crests, but also their
volume is decreasing at high amplitudes.

We define wave energy~$E$ as the excess of mechanical energy due to
the presence of the wave:
\begin{equation}
  \begin{split}
    E = \frac g 2 \int y^2x_\xi d\xi
    - \frac{1}{2} \int \varphi \hat{H}\varphi_\xi d\xi \\
    + \frac{\omega^2}{6} \int y^3 x_\xi d\xi
    + \omega \int \varphi yy_\xi d\xi.
  \end{split}
\end{equation}
It is shown in figure \ref{fig:stat_prm:energy} as a function of wave
height for various values of wave period. As the figure shows, energy
attains a maximum at some critical height, which depends on the
wavelength $L$. Waves of greater height are unstable --- this is in
contrast with BO solitons, which are stable for all amplitudes. The
associated superharmonic instability was studied for waves on linear
shear current by Murashige and Choi \cite{murashige2020stability}.
Positions of energy and volume extrema as functions of wave lenght $L$
are shown in figure \ref{fig:stat_prm:maxima}; as one can see, the
critical height changes very little with increase of $L$, approaching
$h\approx 0.467\lambda_g$ for long waves.

\section{\label{sec:sim}Numerical simulations}

In this section we study dynamic behaviour of solitary waves,
including their formation from an initial perturbation and interaction
of two solitary waves. To model the dynamics of the system
(\ref{eq:kin_bc})--(\ref{eq:inf_bc}), we integrate numerically the
governing equations in conformal coordinates
(\ref{eq:rt}) and (\ref{eq:vt}) with periodic boundary conditions using
the fifth order Dormand--Prince method of the Runge-Kutta family with a
built-in error estimate.

\subsection{\label{sec:pulses}Disintegration of a localized perturbation}

\begin{figure}
  \includegraphics[width=0.9\linewidth]{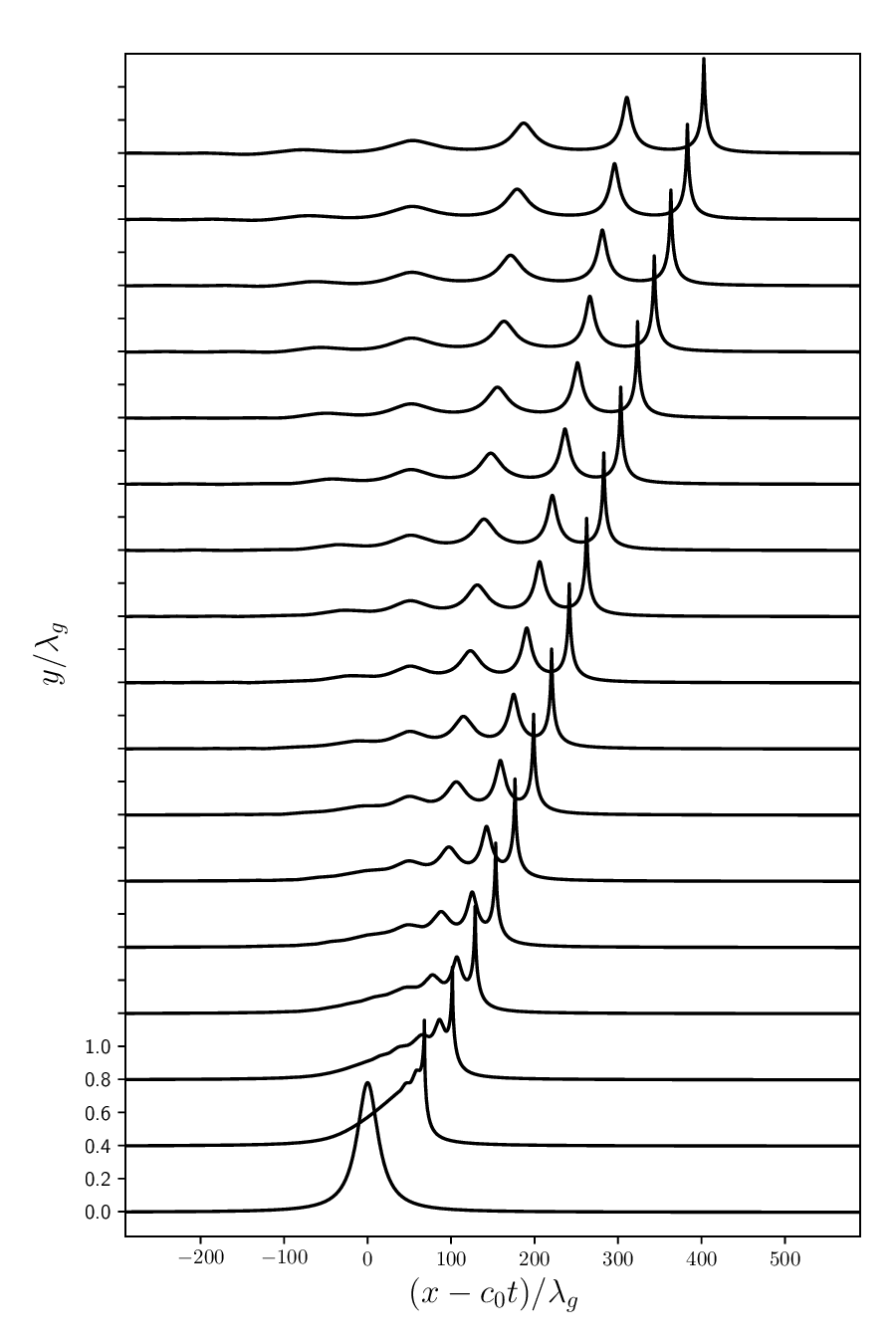}
  \caption{Disintegration of a Lorentz pulse with height
    $a=0.78\lambda_g$ and width $d=15.3\lambda_g$, and the subsequent
    formation of three solitary waves; global period
    $L=10^4\lambda_g$, time step between snapshots
    $|\omega|\Delta t=186$, lower profiles correspond to earlier
    stages.}
  \label{one_to_three}
\end{figure}

\begin{figure}
  \includegraphics[width=\linewidth]{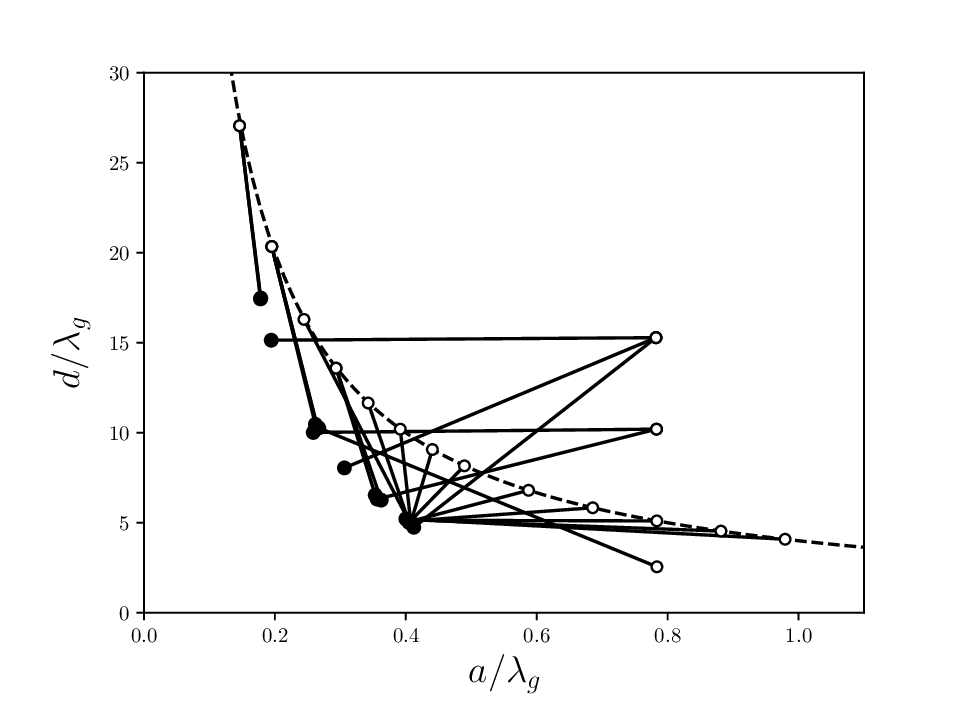}
  \caption{Parameters of initial Lorentzian pulses ($a$ and $d$, white
    markers) and characteristics of solitary waves formed as a result
    of their disintegration (height $h$ and half-width at half-maximum
    $d$, black markers). Edges show the relation between the initial
    conditions and the products. Dashed line corresponds to equation
    (\ref{eq:bo_ad}).}
  \label{edges}
\end{figure}

\begin{figure}
  \centering
  \begin{subfigure}{\linewidth}
    \includegraphics[width=\linewidth]{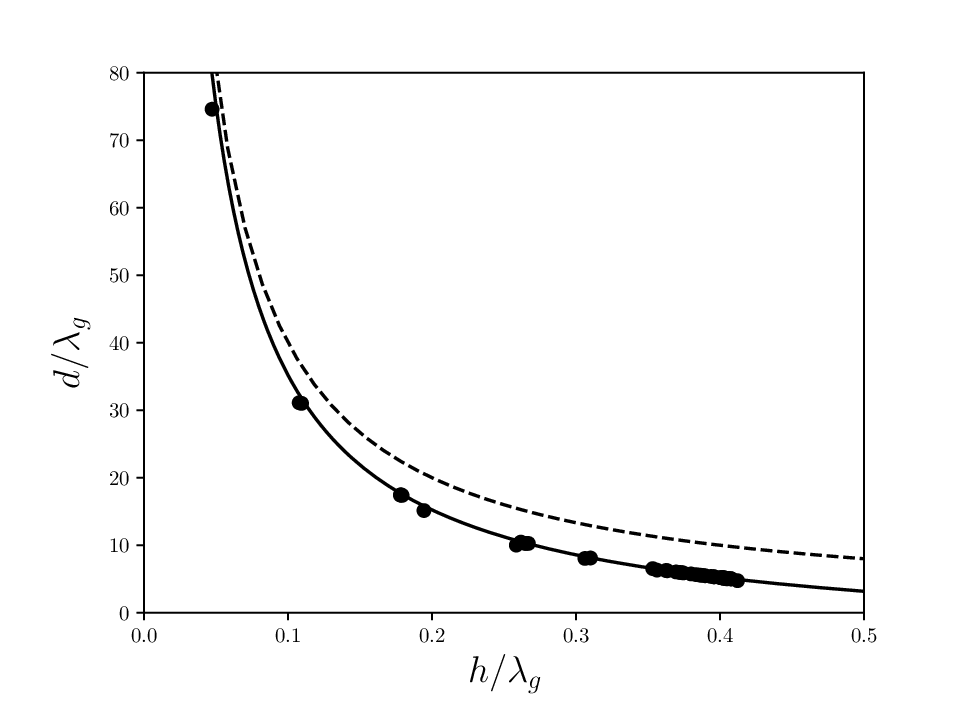}
    \caption{}
    \label{fig:sol_prm:a_d}
  \end{subfigure}
  \begin{subfigure}{\linewidth}
    \includegraphics[width=\linewidth]{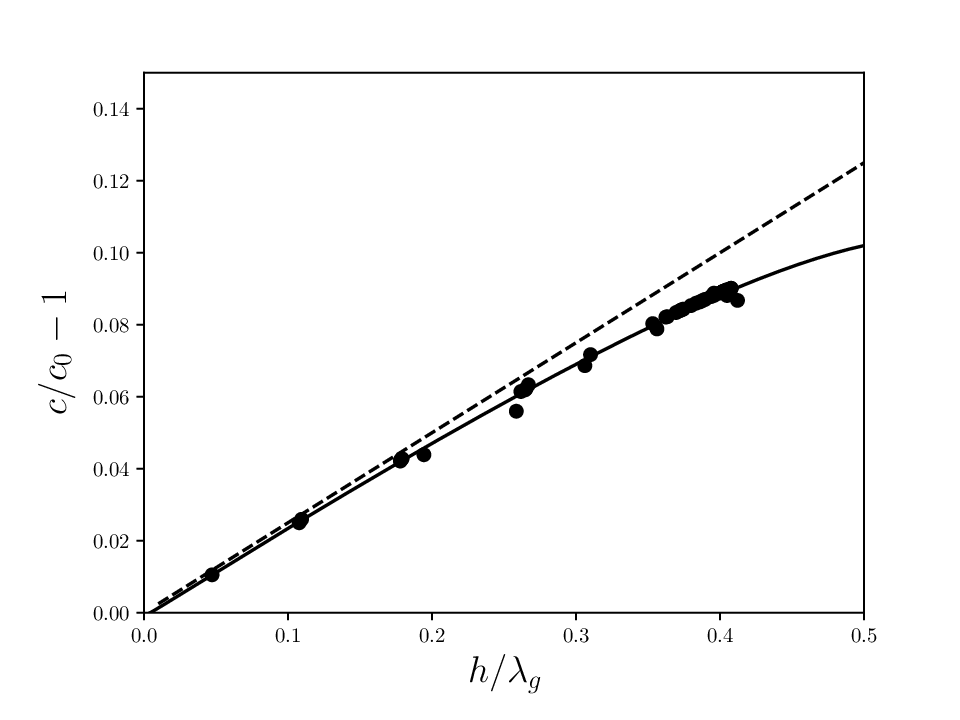}
    \caption{}
    \label{fig:sol_prm:a_c}
  \end{subfigure}
  \caption{Characteristics of solitary waves vs wave height $h$:
    (\subref{fig:sol_prm:a_d})~half-width at half-maximum;
    (\subref{fig:sol_prm:a_c})~phase velocity. Markers correspond to
    solitary waves formed as a result of initial disturbance
    disintegration, solid lines --- to solutions of exact equations
    (with the same spatial period $L=10^4\lambda_g$), dashed lines ---
    to BO solitons.}
  \label{fig:sol_prm}
\end{figure}

In order to observe solitary wave formation from an initial
perturbation we prescribed the initial conditions in a form of a
Lorentzian pulse
\begin{equation}\label{prt_prof}
  f(x)=\frac{a}{1+x^2/d^2}W(x),\qquad W(x)=\cos^2 \frac{\pi x}{L},
\end{equation}
where a window function $W(x)$ was introduced to smooth a jump of
derivatives at the ends of the global period. Height~$a$ and
half-width at half-maximum~$d$ of the pulse were changed
independently, without being necessarily bound by~(\ref{eq:bo_ad}).
The velocity profile was prescribed according
to~(\ref{eq:stat_kin_bc}), with the phase velocity~$c$ computed
using~(\ref{eq:bo_c}) for the soliton of the same height~$a$.  In all
experiments, the global spatial period~$L=10^4\lambda_g$ was at least
$10^2$~times greater than the initial pulse width~$d$.

As a result of disintegration of pulses~(\ref{prt_prof}) with various
parameters, single or multiple solitary waves, as well as packages of
oscillatory waves were typically formed. An example in
figure~\ref{one_to_three} shows how the disintegration of a pulse with
width more than three times greater than that of a BO soliton of the
same height leads to formation of multiple solitons.
Figure~\ref{edges} shows how parameters of generated waves depend on
parameters of initial disturbance. A noteworthy feature of these
results is that for a wide range of initial heights the disintegration
products of pulses, whose parameters \emph{do}
satisfy~(\ref{eq:bo_ad}), seem to accumulate in the vicinity of
height~$0.42\lambda_g$ (which is also slightly less than the critical
height). We did not conduct a parameter space study detailed enough to
tell whether this is a general behaviour for a wider class of initial
conditions. In figure~\ref{fig:sol_prm} half-widths at half-maximum
$d$ of the formed solitary waves and their phase velocities $c$ are
plotted against wave height; as the figure shows, parameters of
generated waves are in reasonable agreement with with the parameters
of stationary waves, obtained in Section~\ref{sec:stwave}.

\subsection{\label{sec:collisions}Solitary wave collisions}

\begin{figure}
  \centering
  \begin{subfigure}{\linewidth}
    \includegraphics[width=\linewidth]{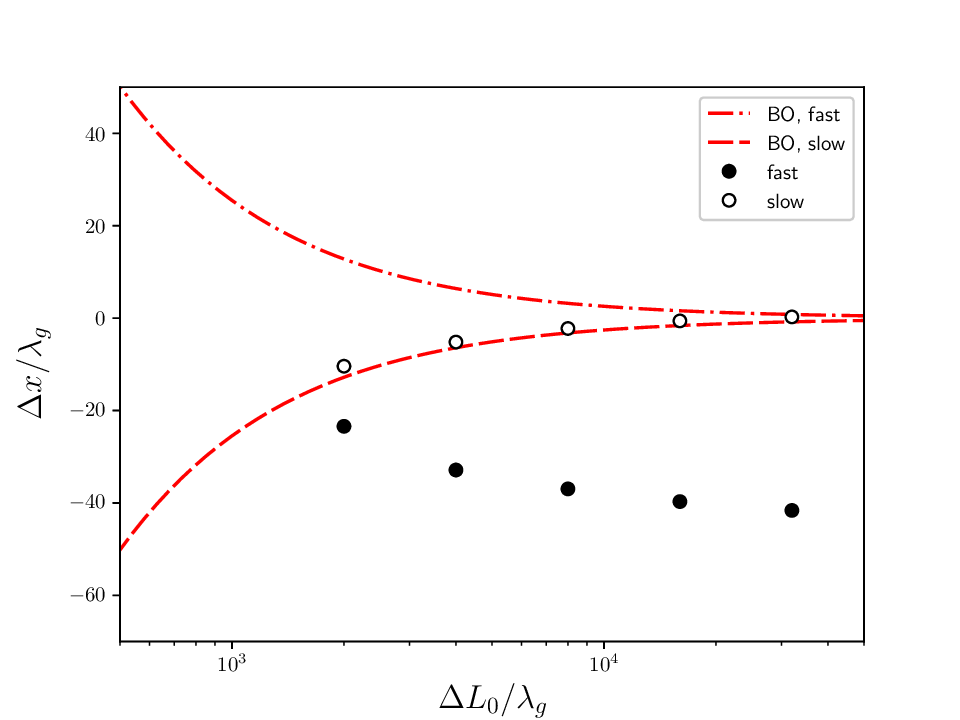}
    \caption{}
    \label{fig:shifts:1}
  \end{subfigure}
  \begin{subfigure}{\linewidth}
    \includegraphics[width=\linewidth]{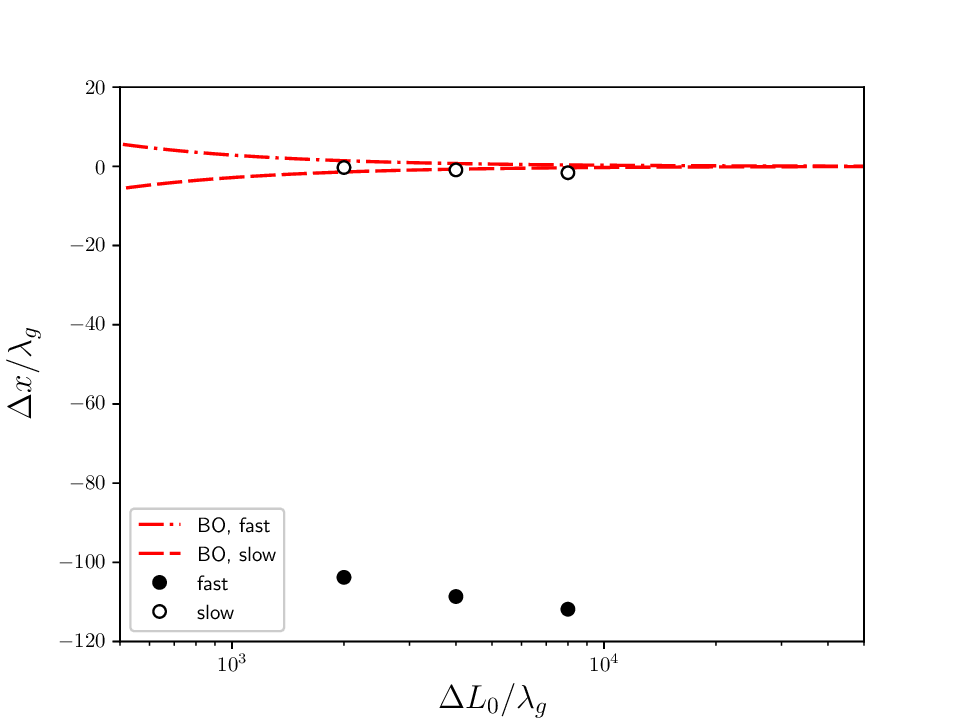}
    \caption{}
    \label{fig:shifts:2}
  \end{subfigure}
  \begin{subfigure}{\linewidth}
    \includegraphics[width=\linewidth]{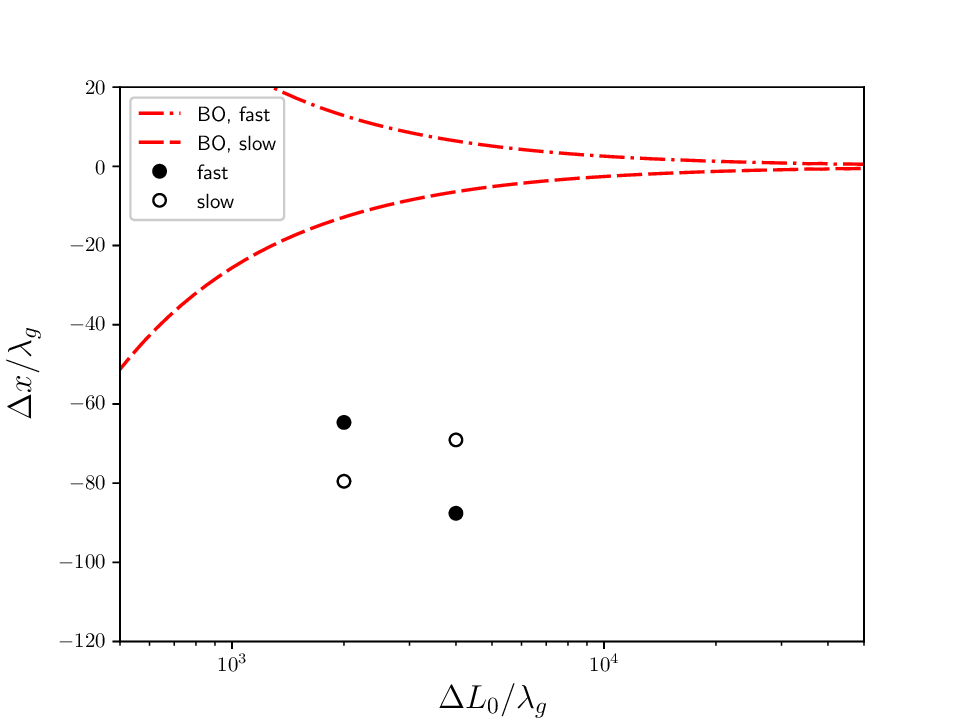}
    \caption{}
    \label{fig:shifts:3}
  \end{subfigure}
  \caption{Phase shifts of solitary waves vs initial distance. The
    parameters: (\subref{fig:shifts:1})~$h_1=0.15\lambda_g$,
    $h_2=0.05\lambda_g$, (\subref{fig:shifts:2})~$h_1=0.40\lambda_g$,
    $h_2=0.10\lambda_g$ and (\subref{fig:shifts:3})~$h_1=0.40\lambda_g$,
    $h_2=0.30\lambda_g$.}
  \label{fig:shifts}
\end{figure}

\begin{figure}
  \centering
  \includegraphics[width=\linewidth]{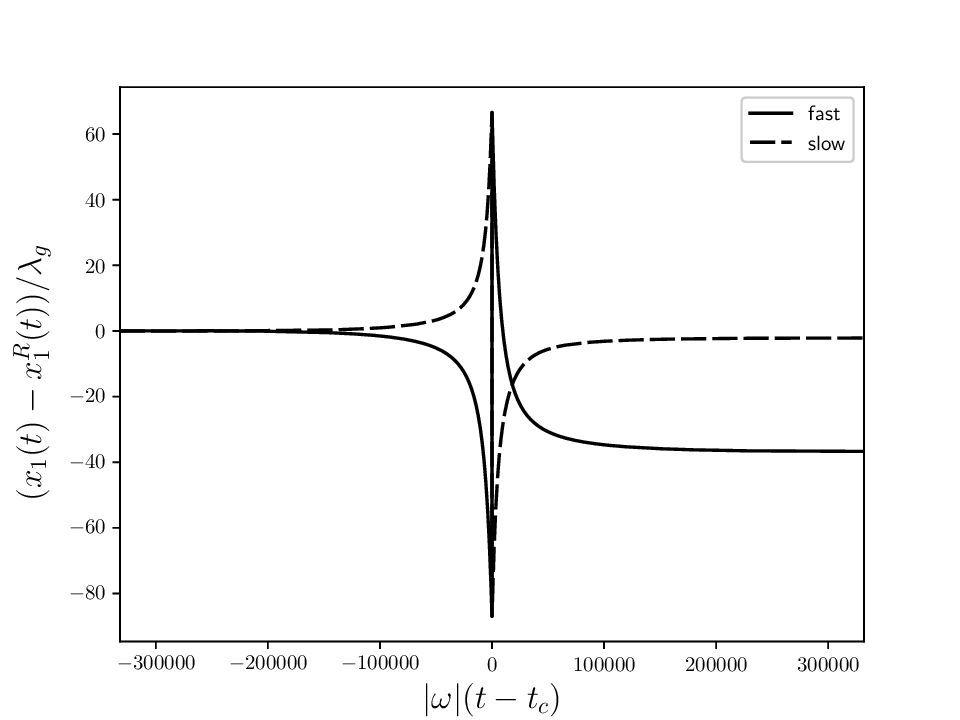}
  \caption{Accumulation of the phase shift of the faster solitary wave
    (solid line) and the slower one (dashed line) during the
    interaction. The parameters: $h_1=0.15$, $h_2=0.05$,
    $\delta L_0=8000\lambda_g$, $L=50\cdot 10^3\lambda_g$}
  \label{fig:accumulation}
\end{figure}

\begin{figure}
  \includegraphics[width=0.9\linewidth]{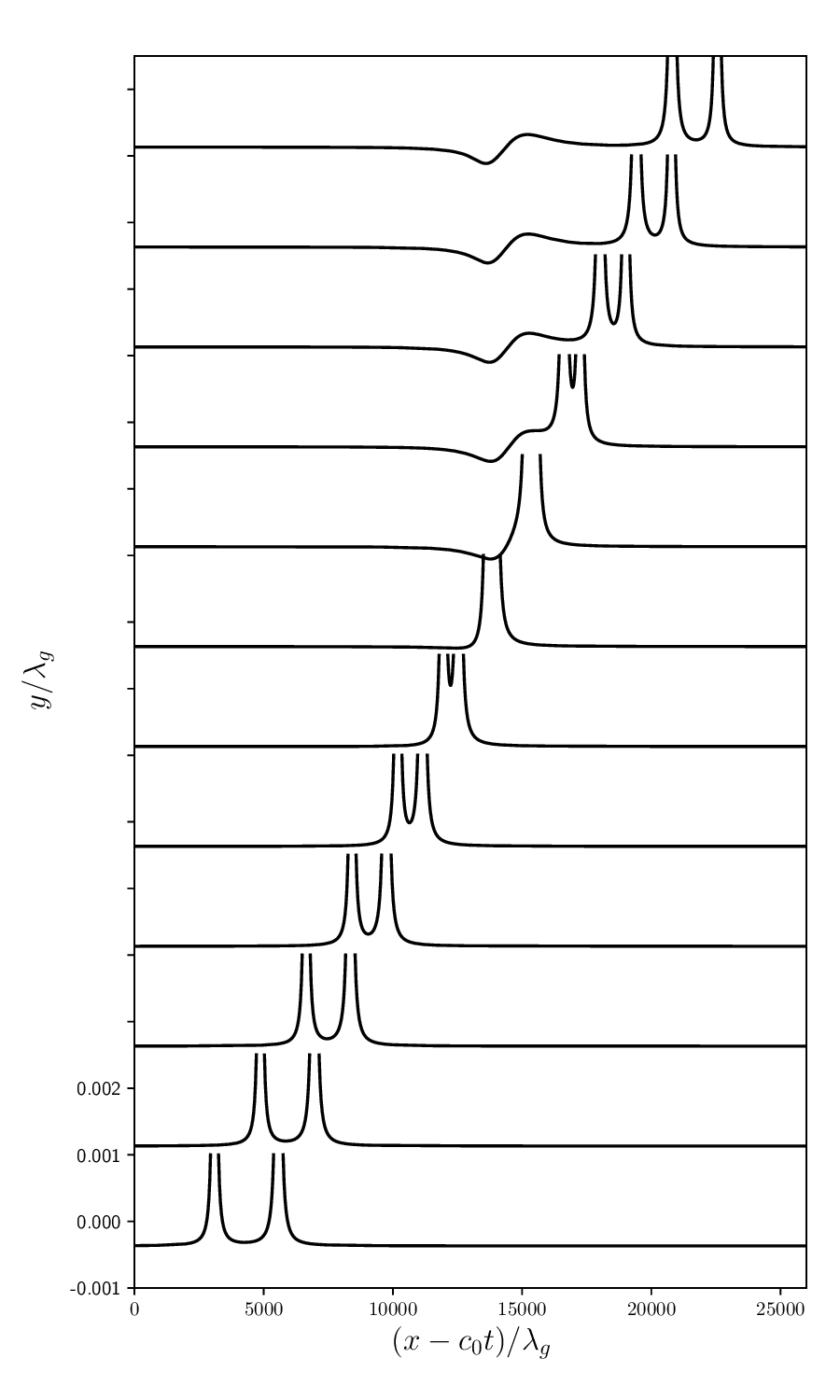}
  \caption{Generation of a step-like perturbation on the free surface
    after a collision of two solitons. The parameters:
    $h_1=0.4\lambda_g$, $h_2=0.3\lambda_g$, and
    $\Delta L_0=4000\lambda_g$, time step between snapshots
    $|\omega|\Delta t=2\cdot 10^4$, lower profiles correspond to
    earlier stages.}
  \label{fig:step}
\end{figure}

\begin{table*}[t]
  \centering
  \begin{tabular}{|c|c|c|c|c|c|c|c|c|}
    \hline
    $h_1$ & $h_2$ & $\Delta L_0$ & $L$ & $\Delta L_{min}$
    & $\Delta x_1$ & $\Delta x_2$ & $h_1'$ & $h_2'$ \\
    \hline
0.1501 & 0.0500 & $2 \cdot 10^3$ & $50 \cdot 10^3$ & 151.97 & -23.41 & -10.40 & 0.1501 & 0.0497 \\
0.1500 & 0.0500 & $4 \cdot 10^3$ & $50 \cdot 10^3$ & 152.84 & -32.87 & -5.18 & 0.1500 & 0.0499 \\
0.1500 & 0.0500 & $8 \cdot 10^3$ & $50 \cdot 10^3$ & 153.06 & -36.96 & -2.24 & 0.1500 & 0.0499 \\
0.1500 & 0.0500 & $16 \cdot 10^3$ & $100 \cdot 10^3$ & 153.11 & -39.70 & -0.60 & 0.1500 & 0.0499 \\
0.1500 & 0.0500 & $32 \cdot 10^3$ & $100 \cdot 10^3$ & 153.13 & -41.62 & 0.27 & 0.1500 & 0.0499 \\
\hline
0.4000 & 0.1000 & $2 \cdot 10^3$ & $20 \cdot 10^3$ & 58.48 & -103.79 & -0.30 & 0.4027 & 0.0966 \\
0.4000 & 0.1000 & $4 \cdot 10^3$ & $20 \cdot 10^3$ & 58.56 & -108.63 & -0.90 & 0.4028 & 0.0978 \\
0.4000 & 0.1000 & $8 \cdot 10^3$ & $50 \cdot 10^3$ & 58.58 & -111.84 & -1.61 & 0.4028 & 0.0980 \\
\hline
0.4000 & 0.3000 & $2 \cdot 10^3$ & $50 \cdot 10^3$ & 242.42 & -64.67 & -79.53 & 0.4004 & 0.2998 \\
0.4000 & 0.3000 & $4 \cdot 10^3$ & $50 \cdot 10^3$ & 243.87 & -87.61 & -69.08 & 0.4003 & 0.2998 \\
\hline
  \end{tabular}
  \caption{Phase shifts and amplitudes of solitary waves after the
    collision. All values are given in units of $\lambda_g$.}
  \label{tab:collision_data}
\end{table*}

To prepare initial conditions containing two waves at the
prescribed positions $\{x_{0n}\}$ we combine inidividual wave profiles
with phase velocities $\{\tilde{c}_n\}$ as
\begin{equation}\label{eq:two_super}
  \begin{split}
    R(\xi) &= 1+\sum_{n=1,2} \left(R_n(\xi-x_{0n})-1\right),\\
    V(\xi) &= \sum_{n=1,2} V_n(\xi-x_{0n}).
  \end{split}
\end{equation}
where $R_n,V_n$ are the Dyachenko representation (\ref{eq:dyach_def})
of the stationary wave profiles, obtained in Section \ref{sec:stwave}.
Phase velocity of waves with constant vorticity is affected by mean
water level; in our setup, where the mean level is kept zero both for
periodic stationary waves and for the superposition
(\ref{eq:two_super}), the actual observed phase velocities of the
crests in the superposition will differ slightly from the original
values $\{\tilde{c}_n\}$, and will be approximately equal to
$c_{1,2} \approx \tilde{c}_{1,2} + \omega S_{2,1}/L$, where $S_{1,2}$ are
volumes of the wave profiles.

Since individual stationary waves have slowly decreasing power-law
tails, it is infeasible within our model to start the simulation from
a point, where interaction between the waves is negligible. We must
therefore study how the result of the interaction depends on initial
distance between the crests $\Delta L_0=x_{02}-x_{01}$.

We are considering only waves propagating in the direction of the
shear, therefore a collision can only occur when a higher and faster
solitary wave catches up with a lower and slower one. Hereinafter we
will use index $n=1$ to refer to the higher solitary wave, and $n=2$
--- to the lower wave. We examine three sets of wave heights: 1) two
low solitary waves: $h_1=0.15\lambda_g,\ h_2=0.05\lambda_g$, 2) one
wave is high, and the other is low:
$h_1=0.40\lambda_g,\ h_2=0.10\lambda_g$, and 3) two high waves:
$h_1=0.40\lambda_g,\ h_2=0.30\lambda_g$. For each pair of wave heights
a number of simulations were performed with varying global period $L$
and initial distance $\Delta L_0$. Each simulation started at $t=0$
with the initial conditions (\ref{eq:two_super}), and continued until
$t=2t_c$, where $t_c=\Delta L_0 / (c_1-c_2)$ is the time when the
collision is expected to occur. We required that $L$ and $\Delta L_0$
satisfy
\begin{displaymath}
  \frac{2\Delta L_0}{c_1-c_2} < \frac{L}{c_1 - c_0}
\end{displaymath}
so that until the end of simulation (i.e., until the two waves are
again separated by the initial distance $\Delta L_0$) the faster wave
would not catch up with any artifacts that were excited at the moment
$t=0$ by the ``artificial'' superposition (\ref{eq:two_super}).

The results of the numerical simulations, which include parameters of
the setup (global period $L$ and initial distance between the waves
$\Delta L_0$), amplitudes of the waves before ($h_{1,2}$ at $t=0$) and
after the collision ($h_{1,2}^\prime$ at $t=2t_c$), minimal distance
between the waves during the interaction $\Delta L_{min}$ and phase
shifts $\Delta x_{1,2}$, are given in table \ref{tab:collision_data}.

All three mentioned pairs of wave heights turned out to demonstrate an
exchange scenario of interaction. When amplitude of the second wave is
sufficiently small, an overtaking also becomes possible; we observed
this behavior, for example, for a pair $h_1=0.35\lambda_g$ and
$h_2=0.02\lambda_g$. We did not study overtaking scenario in greater
detail.

The method of measuring the phase shifts requires special attention,
because even small discrepancy in the involved values of phase
velocity can lead, over long time intervals, to a large inaccuracy in
the resulting phase shift values. We also need to take into account
the possibility for the phase velocity after the collision to be
slightly different from the initial velocity, due to the interaction
being inelastic. We therefore adopted the following definition for the
phase shift, based on the actual velocity of the solitary wave crest
\begin{equation}\label{eq:shift_def}
  \Delta x_n = \left[x_n(2t_c) - x_n(0)\right]
  - \left[\dot{x}_n(0) + \dot{x}_n(2t_c)\right]t_c
\end{equation}
where $x_n(t)$ is the observed trajectory of the n-th wave (defined as
the position of the local elevation maximum), $\dot x_n(t)$ --- its
velocity. This is equivalent to measuring position of the crest
relative to the reference trajectory
\begin{equation}\label{eq:rtraj_def}
  x_n^R(t) = \begin{cases}
    x_n(0) + \dot{x}_n(0)t,\ &t \leq t_c\\
    x_n(0) + \dot{x}_n(0)t_c + \dot{x}_n(2t_c)(t-t_c),\ &t > t_c
  \end{cases}
\end{equation}
that is, $\Delta x_n=x_n(2t_c)-x_n^R(2t_c)$.

The phase shift definition (\ref{eq:shift_def}), which uses only a
finite piece of the wave trajectory in the vicinity of the collision
point, would actually produce non-zero values even for the two-soliton
solution of the BO equation (although the ``apparent'' phase shifts of
the BO solitons vanish at large observation times). This is mostly due
to the fact, that the faster soliton ``jumps'' forward during the
exchange (and the slower one ``jumps'' backward). In figure
\ref{fig:shifts} the measured phase shifts in the numerical
simulations are plotted against the initial distance $\Delta L_0$;
apparent phase shifts for the two-soliton BO solution
\cite{matsuno1980interaction}, computed as in (\ref{eq:shift_def}) for
the same pair of soliton amplitudes and the same initial distance
$\Delta L_0$, are plotted as a reference. As the figure shows, in the
first (low+low pair) and the second (high+low) series of simulations
phase shifts of the slower solitary waves closely resemble the
behaviour of the apparent phase shifts of the slower solitons in the
BO two-soliton solution, and approach (for low+low) or stay in the
vicinity of zero (for high+low) as $\Delta L_0$ increases. Meanwhile,
phase shifts of the faster waves grow in absolute value, approaching
constant negative levels. Figure \ref{fig:accumulation} shows a
typical example of how the position of the faster solitonary wave
relative to the reference trajectory (\ref{eq:rtraj_def}) changes over
time; one can see, that the phase shift is developed shortly after the
collision, remaining almost constant until the end of the simulation.

Unfortunately, we were not able to conduct many experiments for the
third pair of solitary waves (high+high) due to the simulation cost
being much greater for that pair than for the first two. Still, from
the available points in figure \ref{fig:shifts:3} it is already clear
that in this pair the behaviour of the slower wave departs from the
predictions of the BO equation as prominently as the behaviour of the
faster one does.

Finally, figure \ref{fig:step} shows generation of a small-amplitude
step-like perturbation on the free surface after the collision of a
pair of high solitons, which indicates the inelastic nature of the
solitary waves interaction. This feature is absent in the two-soliton
solution of the BO equation, which at all times is represented by a
sum of two poles. However, as indicated in the Table
\ref{tab:collision_data}, in all our simulations the wave amplitudes
after the collision remain close to the initial amplitudes, from which
we can conclude that the interaction is almost elastic.

\section{\label{sec:conclusion}Conclusions}

The existence of solitons on deep water with constant vorticity
propagating in the direction of the shear was demonstrated in earlier
research (see Shrira \cite{shrira1986nonlinear}) in the limit of large
wavelengths (i.e., weak dispersion) and small amplitudes. It was
unknown whether the solitons retain their properties with decrease in
the characteristic wavelength and transition to greater
amplitudes. The results of the numerical simulations presented in this
paper show that the fully nonlinear equations of motion allow the
existence of stable solitary waves as well, although the waves
exceeding critical wave height become unstable. Numerical simulations
of solitary wave collisions provide a reliable evidence that within
the exact equations of motion, unlike the Benjamin-Ono equation, waves
undergo a phase shift as a result of the interaction. Despite the fact
that generation of oscillatory waves was observed during the
collision, solitary wave amplitudes remained close to their initial
values; this may indicate that the interaction is almost elastic and
an approximate integrable model of this process may exist outside the
range of applicability of the Benjamin-Ono equation.

In our study, we assumed deep water regime. If we consider long waves
of length $L$ with localized crests in fluid of finite depth $D$, the
deep water regime for them would imply that the intrinsic length scale
$\lambda_g$ due to vorticity $\omega$ has to be much smaller than the
water depth, i.e.,  \mbox{$\lambda_g = g/\omega^2 \ll L < 2D$}. The
magnitude of vorticity must be large enough to provide sufficient
separation of scales between $\lambda_g$ and $D$. Here, we take into
account the fact that, as we observed in our study, for long waves
there exists a critical wave height and, therefore, a minimum crest
width of the order of a few $\lambda_g$; the waves must be much longer
than $\lambda_g$ for their crests to be localized in a small fraction
of wavelength. The condition $\lambda_g \ll D$ may be difficult to
satisfy for the water waves, as it would require unrealistically large
values of vorticity. An example of a geophysical process, where this
requirement is easier to satisfy, is internal waves propagating
zonally on equator, where constant vorticity is provided by the Earth
rotation, while the effective gravitational acceleration $g^*$ is
greatly reduced. In context of geophysical applications our
simulations can be viewed as a toy model aimed at understanding of
this very special class of wave motions.

\section*{Acknowledgements}

The authors would like to express their gratitude to V.~I.~Shrira and
K.~A.~Gorshkov for providing valuable suggestions. This research was
supported by the Russian Foundation for Basic Research (Grant No.
21-55-52005 MNT\_a), Russian Science Foundation (Grant No. 19-17-00209)
and by the President of the Russian Federation (Grant No. MK-2041.2017.5).

\section*{Data availability}

The data that support the findings of this study are available within
the article.

\appendix

\section{Derivation of the Benjamin-Ono equation}

We use the system of equations \cite{shishina2016}
\begin{subequations}\label{r3_12}
  \begin{equation}\label{r3_12:a}
    y_t(1+\tilde x_\xi) - x_ty_\xi - \omega yy_\xi = -\hat{H}\varphi_\xi,
  \end{equation}
  \begin{equation}\label{r3_12:b}
    \begin{split}
      \varphi_ty_\xi - &\varphi_\xi y_t + gyy_\xi + \omega \psi y_\xi \\
      + \hat{H}\big\{&
        \varphi_t(1+\tilde{x}_\xi) - \varphi_\xi x_t + gy(1+\tilde{x}_\xi)\\
      &+ \omega(\psi(1+\tilde{x}_\xi)-\varphi_\xi y)
        \big\} = 0.
     \end{split}
  \end{equation}
\end{subequations}
Let us consider the linearized problem:
\begin{eqnarray*}
  &&y_t = -\hat{H}\varphi_\xi,\\
  &&gy + \varphi_t + \omega \hat{H}\varphi = 0.
\end{eqnarray*}
Hence we arrive at the integrodifferential equation
\begin{displaymath}
  \varphi_{tt} + \omega\hat{H}\varphi_t -g\hat{H}\varphi_\xi = 0.
\end{displaymath}
From the dispersion relation (\ref{eq:lin_disp}), we find the linear
dispersion law:
\begin{displaymath}
  \Omega_{1,2} = \frac{\omega}{2} \pm \sqrt{\frac{\omega^2}{4}+gk}.
\end{displaymath}

At $gk \ll \omega^2$ for long, weakly dispersive waves propagating
against the flow ($\omega < 0$), the linear dispersion relation
\begin{displaymath}
  \Omega = \frac{\omega}{2} - \frac{\omega}{2}\sqrt{1+\frac{4gk}{\omega^2}}
\end{displaymath}
goes to
\begin{equation}\label{eq:disp_law}
  \Omega = -\frac{gk}{\omega} + \frac{g^2k^2}{\omega^3}.
\end{equation}
(The value $c_0=-\frac{g}{\omega}$ by the dimension represents
the velocity of long waves with small amplitude).

Let us show that the evolution of nonlinear waves corresponding to the
linear dispersion law (\ref{eq:disp_law}) is described by the
Benjamin-Ono equation, whose solitons have the form of a Lorentz
pulse. In the system of equations (\ref{r3_12}) we make
the substitution
\begin{displaymath}
  y(\xi,t) \to y(\zeta,t), \quad \varphi(\xi,t) \to \varphi(\zeta,t),
\end{displaymath}
where
\begin{equation}\label{subst}
  \zeta = \varepsilon \left(\xi + \frac{g}{\omega}t\right),
  \quad
  \tau = -\varepsilon^2\frac{g}{\omega}t,
\end{equation}
and where $\varepsilon$ is a small parameter.

Substituting (\ref{subst}) into the system of equations (\ref{r3_12}),
after some algebra, we have
\begin{subequations}
  \label{r53}
  \begin{equation}\label{r53:a}
    \frac{g}{\omega}y_\zeta - \varepsilon\frac{g}{\omega} y_\tau (1+\varepsilon \tilde{x}_\zeta)
    + \varepsilon^2\frac{g}{\omega}\tilde{x}_\tau y_\zeta - \omega yy_\zeta = -\hat{H}\varphi_\zeta
  \end{equation}
  \begin{eqnarray}\label{r53:b}
    &&gy(1+\varepsilon \tilde{x}_\zeta) + \varepsilon\frac{g}{\omega}\varphi_\zeta
       - \varepsilon^2\frac{g}{\omega}\varphi_\tau(1+\varepsilon \tilde{x}_\zeta)\nonumber\\
    &&+ \varepsilon^3\frac{g}{\omega}\varphi_\zeta\tilde{x}_\tau
       - \varepsilon\omega y\varphi_\zeta + \omega\psi(1+\varepsilon \tilde{x}_\zeta) \nonumber\\
    &&=\hat{H}(
       \varepsilon^3\frac{g}{\omega}y_\tau\varphi_\zeta - \varepsilon^3\frac{g}{\omega}y_\zeta\varphi_\tau
       + \varepsilon gyy_\zeta + \varepsilon\omega\psi y_\zeta
       )
  \end{eqnarray}
\end{subequations}
We seek the solution of the system (\ref{r53}) in the form of a
series with respect to $\varepsilon$:
\begin{eqnarray}
  \label{r54}
  y(\zeta,\tau)&=&\varepsilon y_1(\zeta,\tau)+\varepsilon^2 y_2(\zeta,\tau)+\ldots\\
  \label{r55}
  \varphi(\zeta,\tau)&=&\varepsilon\varphi_1(\zeta,\tau)+\varepsilon^2\varphi_2(\zeta,\tau)+\ldots
\end{eqnarray}
We substitute (\ref{r54}) and (\ref{r55}) into system (\ref{r53}) and
select the terms of the same orders of magnitude. In the first
approximation
\begin{eqnarray*}
  &&\frac{g}{\omega}\frac{\partial y_1}{\partial\zeta} = -\hat{H}\frac{\partial\varphi_1}{\partial\zeta}\\
  &&gy_1+\omega\hat{H}\varphi_1=0.\nonumber
\end{eqnarray*}
In the following order
\begin{subequations}
  \label{r57}
  \begin{equation}
    \label{subeq:a}
    \frac{g}{\omega}\frac{\partial y_2}{\partial\zeta}
    -\frac{g}{\omega}\frac{\partial y_1}{\partial\tau}
    -\omega y_1\frac{\partial y_1}{\partial\zeta} = -\hat{H}\frac{\partial\varphi_2}{\partial\zeta},
  \end{equation}
  \begin{equation}
    \label{subeq:b}
    gy_2 + \frac{g}{\omega}\frac{\partial\varphi_1}{\partial\zeta}
    + \omega\hat{H}\varphi_2=0.
  \end{equation}
\end{subequations}
System (\ref{r57}) leads to the differential equation
\begin{displaymath}
  \frac{g}{\omega}\frac{\partial y_1}{\partial\tau}
  + \omega y_1\frac{\partial y_1}{\partial\zeta}
  + \frac{g}{\omega^2}\frac{\partial^2\varphi_1}{\partial\zeta^2}=0.
\end{displaymath}
Considering that
\begin{displaymath}
  \frac{\partial^2\varphi_1}{\partial\zeta^2}=
  \frac{g}{\omega}\hat{H}\frac{\partial^2y_1}{\partial\zeta^2},
\end{displaymath}
we obtain the following Benjamin-Ono equation ($y_1=y$) (omitting index 1):
\begin{equation}
  \label{r58}
  \frac{g}{\omega}\frac{\partial y}{\partial\tau}
  + \omega y\frac{\partial y}{\partial\zeta}
  + \frac{g^2}{\omega^3}\hat{H}\frac{\partial^2y}{\partial\zeta^2}=0.
\end{equation}
Using the initial variables, the equation (\ref{r58}) takes the form of
the equation (\ref{eq:bo_eqn}).

\nocite{*}
\bibliography{solitons}

\end{document}